# Joint CSI Estimation, Beamforming and Scheduling Design for Wideband Massive MIMO System

Tadilo Endeshaw Bogale$^{+}$, Long Bao Le$^{+}$, and Xianbin Wang$^{++}$
Institute National de la Recherche Scientifique (INRS)
Université du Québec, Montréal, Canada$^{+}$
University of Western Ontario, London, Canada$^{++}$
Email: {tadilo.bogale, long.le}@emt.inrs.ca, xianbin.wang@uwo.ca

***Abstract*— This paper proposes a novel approach for designing channel estimation, beamforming and scheduling jointly for wideband massive multiple input multiple output (MIMO) systems. With the proposed approach, we first quantify the maximum number of user equipments (UEs) that can send pilots which may or may not be orthogonal. Specifically, when the channel has a maximum of $L$ multipath taps, and we allocate $\tilde{M}$ sub-carriers for the channel state information (CSI) estimation, a maximum of $\tilde{M}$ UEs CSI can be estimated ($L$ times compared to the conventional CSI estimation approach) in a massive MIMO regime. Then, we propose to schedule a subset of these UEs using greedy based scheduling to transmit their data on each sub-carrier with the proposed joint beamforming and scheduling design. We employ the well known maximum ratio combiner (MRC) beamforming approach in the uplink channel data transmission. All the analytical expressions are validated via numerical results, and the superiority of the proposed design over the conventional orthogonal frequency division multiplexing (OFDM) transmission approach is demonstrated using extensive numerical simulations in the long term evolution (LTE) channel environment. The proposed channel estimation and beamforming design is linear and simple to implement.**

***Index Terms*— Beamforming, Channel estimation, Massive MIMO, Multipath taps, OFDM, Pilot design, Scheduling**

## I. Introduction

Massive multiple input multiple output (MIMO) technology is one of the promising means for achieving the energy and spectrum efficiency requirements of the future 5G networks [1]–[3]. The beamforming gain envisaged by the massive MIMO system depends on the availability of channel state information (CSI). In general, the channel between the transmitter and receiver is estimated from orthogonal pilot sequences where the number of such sequences is limited by the channel coherence time [4]–[7]. In a multiuser setup, massive MIMO can be deployed both at the base stations (BSs) and mobile stations (UEs). However, due to space, energy consumption and cost constraints, deploying massive number of antennas at the UEs is usually infeasible from practical viewpoint particularly at microwave frequency bands. For this reason, it is economical to deploy antenna arrays at the BSs and single antennas at each UE for microwave frequency band applications. This motivates us to consider a massive MIMO system where each UE has single antenna as in [4].

In a massive MIMO system, channel estimation can be performed either in frequency division duplex (FDD) or time division duplex (TDD) approach. In both approaches, the pilot sequences need to be orthogonal to maintain sufficient channel estimation quality. In a traditional pilot transmission, the number of pilots may need to be the same as the number of transmitted antennas to maintain orthogonality. Thus, for a massive MIMO system, one can estimate the maximum number of channel coefficients using time division duplex (TDD) approach by sending pilots from UEs. By doing so, each BS can estimate and utilize the channel coefficients both for the downlink and uplink transmissions. A number of approaches have been proposed to address channel estimation for massive MIMO systems including a sub-space method which exploits the sub-space information of the channel covariance matrices [5], [7], [8], pilot optimization/scheduling/shifting method which optimizes/schedules/shifts the pilot sequences of each cell (UE) while utilizing appropriate signal dimensions such as time and frequency [6], [9]–[11].

The problem of channel estimation for wideband massive MIMO system is studied in the context of multicell system in [12]. When the channel has $L$ multipath taps, each cell has $\tilde{M}$ sampling periods for pilot transmission and each BS serves the number of UEs equal to $\tilde{K} = \frac{\tilde{M}}{L}$ which is the number of UEs obtained with orthogonal pilots (see (4) of [4] and Section II of [10]), it is shown in [12] that at least $L$ cells can reliably estimate the channels of their user equipments (UEs) and perform beamforming without experiencing pilot contamination. This result has the same significance as increasing the number of UEs by $L$ times for the same time/frequency resource. Although such an increment is useful, the design approach of [12] has meaningful improvement compared to the traditional orthogonal pilot in the massive MIMO regime. The main reason the approach of [12] performs badly in a small number of antennas regime is that in such a regime, each UE will experience strong interference signal from the other UEs which are generally large compared to the one used in orthogonal pilots. This motivates us to schedule the number of UEs adaptively while maximizing the total sum spectrum efficiency (SE). This leads to the joint design of CSI estimation, scheduling and beamforming for wideband massive MIMO system which is the focus of the current paper. In [12], pilot transmission is performed in time domain (i.e., non-orthogonal frequency division multiplexing (OFDM)). However, in the current paper, we apply pilot transmission in frequency domain. As will be clear in the sequel, such a pilot transmission approach naturally leads the scheduling design to select only a subset of the UEs for transmission.

In particular, the current paper proposes a novel approach for designing channel estimation, beamforming and scheduling jointly for wideband massive MIMO systems. With the proposed approach, we first quantify the maximum number of UEs that can send pilots which may or may not be orthogonal. Specifically, when the channel has a maximum of $L$ multipath taps, and allocate $\tilde{M}$ sub-carriers for the CSI estimation, a maximum of $\tilde{M}$ UEs can use these sub-carriers for their CSI estimation (i.e., $L$ times more UEs compared to the traditional minimum mean square error (MMSE) CSI estimation approach) in a massive MIMO regime. Then, we select only a subset of these UEs to transmit their data for each sub-carrier with the proposed joint beamforming and scheduling design. We employ the well known maximum ratio combiner (MRC) beamforming approach in the uplink channel data transmission. All the analytical expressions are validated via numerical results, and the superiority of the proposed design over the conventional transmission approach is demonstrated using extensive numerical simulations for LTE channel models. The proposed channel estimation and beamforming design is linear and simple to implement.

This paper is organized as follows. Section II discusses the system and channel models. In Section III, a brief summary of conventional OFDM based pilot and data transmission is discussed. In Sections IV and V, the proposed joint channel estimation, beamforming and scheduling design is provided. In Section VI, extensive simulation results are presented. Finally, Section VII draws conclusions.

## II. System and Channel Model

We consider a multiuser system with a transmission scheme having a sampling period $T_s$ and a channel having a maximum delay spread $T_d$. It is assumed that the BS and each UE are equipped with $N$ antennas and 1 antenna, respectively. For this setting, the number of multipath channel impulse response (CIR) taps $L$ is approximated as $L = \frac{T_d}{T_s}$ [13]. The multipath coefficients between the $k$th UE and $n$th BS antenna is denoted as [12]

$$\bar{\mathbf{h}}_{kn} = [\bar{h}_{kn1}, \bar{h}_{kn2}, \cdots, \bar{h}_{knL}]^T. \tag{1}$$

The analytical model for $\bar{\mathbf{H}}_k = [\bar{\mathbf{h}}_{k1}, \bar{\mathbf{h}}_{k2}, \cdots, \bar{\mathbf{h}}_{kN}]$ may vary from one standard to another. In this paper, we assume that $\bar{\mathbf{h}}_{kn}$ are correlated temporarily over multipath taps and spatially over the BS antenna arrays. We further assume that the temporal correlation matrices corresponding to all antennas are the same, and the spatial correlation matrix of all multipath components are the same. Under such assumptions, $\bar{\mathbf{H}}_k$ is modeled as [14]

$$\bar{\mathbf{H}}_k = \sqrt{\mathbf{P}}_k \tilde{\mathbf{H}}_k \sqrt{\mathbf{R}}_k \tag{2}$$

where each entry of $\tilde{\mathbf{H}}_k$ is an i.i.d ZMCSCG random variable with unit variance, $\mathbf{P}_k \in \mathcal{C}^{L \times L}$ and $\mathbf{R}_k \in \mathcal{C}^{N \times N}$ are positive semidefinite temporal and spatial correlation matrices, respectively. Furthermore, the correlation between an antenna and itself (i.e., diagonal elements of $\mathbf{R}_k$) is assumed to be 1.

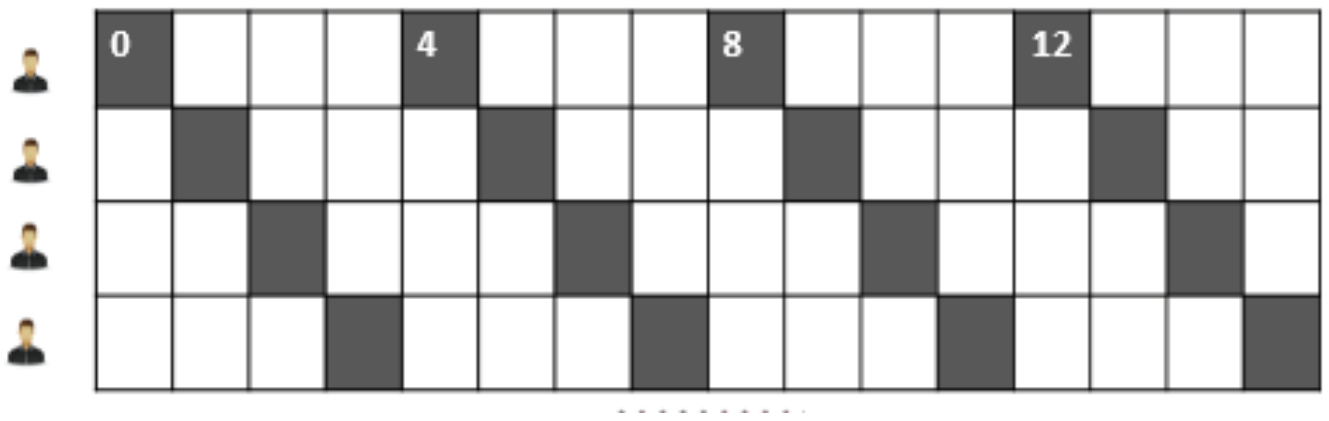

Fig. 1: Existing pilot allocation.

In most wireless standards (e.g., Long term evolution (LTE)), OFDM based transmission is commonly adopted. For such a transmission, the channel coefficient of each sub-carrier has practical importance which can be obtained by linearly combining the multipath tap coefficients (1). To this end, the channel between the $k$th UE and $n$th antenna BS in sub-carrier $s$ can be expressed as [13], [15]

$$h_{kns} = \tilde{\mathbf{f}}_s^H \bar{\mathbf{h}}_{kn} \tag{3}$$

where $\tilde{\mathbf{f}}_s^H = [1, e^{-j\frac{2\pi}{\tilde{M}}s}, e^{-j\frac{2\pi}{\tilde{M}}2s}, \cdots, e^{-j\frac{2\pi}{\tilde{M}}(L-1)s}]$ with $\tilde{M}$ being the fast Fourier transform (FFT) size of the OFDM during pilot transmission. As can be seen from this expression, for fixed $L$, $\bar{\mathbf{h}}_{kn}$ and sampling period $T_s$, the CSI coefficient at sub-carrier $s$ is different for different $\tilde{M}$.

## III. Conventional pilot and data transmission approach

Here we discuss the conventional pilot and data transmission approaches. Such an approach first utilizes orthogonal pilots for each sub-carrier during channel estimation as shown in Fig. 1 for the settings $L = 4$ and $\tilde{M} = 16$. As can be seen from this figure, each UE may need to spend at least $L = 4$ pilots for this setup to maintain orthogonality (For example, the first UE utilizes sub-carriers 0, 4, 8 and 12). This shows that with the conventional approach, the BS can estimate the CSI information for a maximum of $\tilde{K} = \frac{\tilde{M}}{L} = 4$ UEs.

In fact, such an estimator yields the anticipated theoretical beamforming performance during the data transmission stage, and maintains a non-decreasing signal to interference plus noise ratio (SINR) as $N$ increases and leads to unbounded SINR when $N \to \infty$. However, if $\tilde{K} > \frac{\tilde{M}}{L}$ (i.e., a reusing of pilots in one or more sub-carriers), the beamforming performance suffers from an SINR floor even if $N \to \infty$. The reason of this phenomena, the reuse of pilot sequences in the same time (frequency) resource, is termed as *pilot contamination* [4].

We would like to emphasize here that pilot transmission can also be performed without OFDM. With this pilot transmission, one also gets $\tilde{K} = \frac{\tilde{M}}{L} = 4$ (see the discussion after (6) of [12], (4) of [4] and Section II of [10]). Thus, with the conventional pilot transmission, transmitting pilots either with OFDM or without OFDM approach does not bring any difference on the maximum number of supported UEs. In addition, the conventional approach estimates each UE's sub-carrier CSI without taking into account the data transmission strategy. Furthermore, with the conventional approach, one can estimate the CSI of each sub-carrier independently for each antenna. In the following, we discuss the proposed joint channel estimation and data transmission approach.

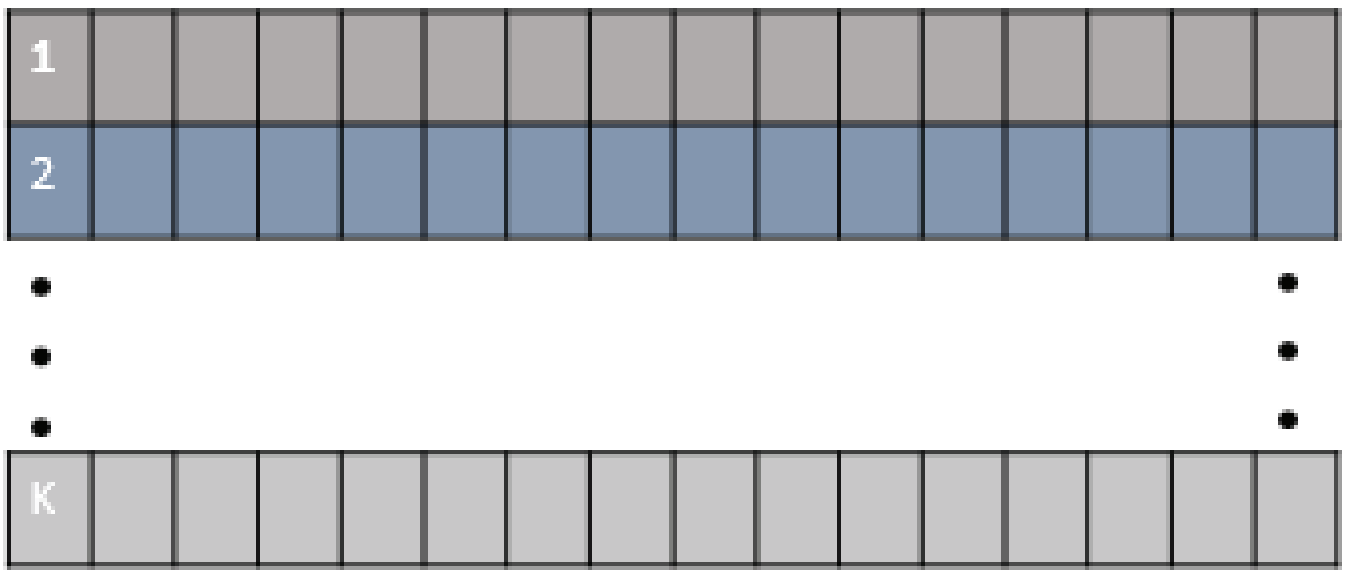

Fig. 2: Proposed pilot allocation.

## IV. Proposed Joint CSI estimation and Beamforming

This section discusses the proposed channel estimation and beamforming approach. We employ the commonly adopted maximum ratio combiner (MRC) beamforming[1].

### A. Proposed Pilot Transmission

The main idea of the proposed pilot transmission is to let every UE to use all the sub-carriers to transmit their pilot symbols as shown in Fig. 2, and then to compute each UE's sub-carrier CSI as a linear combination of the received signals of all sub-carriers, where the linear combination coefficients and pilot sequences are selected carefully to ensure that each UE's sub-carrier SINR increases as $N$ increases (and it grows unbounded as $N \to \infty$)

If we assume the total number of supported UEs as $K$, the received signal by the $n$th antenna at sub-carrier $s$ becomes

$$r_{ns} = \sum_{i=1}^{K} \tilde{\mathbf{f}}_s^H \mathbf{h}_{in} x_{is} + w_{ns}$$

where $x_{is}$ is the pilot symbol sent by the $i$th UE in sub-carrier $s$ and $w_{ns}$ the additive noise at the $n$th BS antenna and $s$th sub-carrier which is assumed to be independent and identically distributed (i.i.d) zero mean circularly symmetric complex Gaussian (ZMCSCG) random variable with variance $\sigma^2$.

The received signals of all antennas and sub-carriers can be expressed as

$$\mathbf{R} = [\mathbf{r}_1, \mathbf{r}_2, \cdots, \mathbf{r}_N] = \sum_{i=1}^{K} \mathbf{X}_i \tilde{\mathbf{F}}^H \mathbf{H}_i + \mathbf{W}$$

where $\mathbf{x}_i = [x_{i1}, x_{i2}, \cdots, x_{i\tilde{M}}]^T$, $\mathbf{r}_n = [r_{n1}, r_{n2}, \cdots, r_{n\tilde{M}}]^T$, $\mathbf{w}_n = [w_{n1}, w_{n2}, \cdots, w_{n\tilde{M}}]$, $\mathbf{X}_i = \mathrm{diag}(\mathbf{x}_i)$, $\mathbf{W} = [\mathbf{w}_1, \mathbf{w}_2, \cdots, \mathbf{w}_N]$, $\mathbf{H}_i = [\mathbf{h}_{i1}, \mathbf{h}_{i2}, \cdots, \mathbf{h}_{iN}]$ and $\tilde{\mathbf{F}} = [\tilde{\mathbf{f}}_1, \tilde{\mathbf{f}}_2, \cdots, \tilde{\mathbf{f}}_{\tilde{M}}]$.

We express the $k$th UE $s$th sub-carrier CSI estimate as

$$\hat{\mathbf{h}}_{ks} = \mathbf{R}^T \mathbf{t}_{ks}. \tag{4}$$

[1]The extension of the proposed technique for other beamforming techniques is discussed in [16].

As we can see from this expression, $\hat{\mathbf{h}}_{ks}$ depends on the linear combination vector $\mathbf{t}_{ks}$ and the pilot matrix $\mathbf{X}_k$ which are designed based on the beamforming strategy during data transmission as discussed below.

### B. Data Transmission

As discussed above, we consider the uplink channel for data transmission using the OFDM approach. For this reason, the received signal of each sub-carrier can be obtained independently at each antenna of all BSs. To this end, the $n$th BS antenna receives the following signal in sub-carrier $s$ ($y_{ins}$)

$$y_{ns} = \sum_{k=1}^{K} \mathbf{f}_s^H \bar{\mathbf{h}}_{kn} d_{ks} + \tilde{w}_{ns}$$

where $\tilde{w}_{ns}$ is the noise sample at the $n$th BS antenna and sub-carrier $s$ during data transmission and $d_{ks}$ is the data symbol transmitted in sub-carrier $s$ of the $k$th UE. The overall received signal at the $s$th sub-carrier becomes

$$\mathbf{y}_s = \sum_{i=1}^{K} \mathbf{H}_i^T \mathbf{f}_s^* d_{is} + \tilde{\mathbf{w}}_s = \mathbf{H}_k^T \mathbf{f}_s^* d_{ks} + \sum_{i \neq k}^{K} \mathbf{H}_i^T \mathbf{f}_s^* d_{is} + \tilde{\mathbf{w}}_s$$

where $\mathbf{y}_s = [y_{1s}, y_{2s}, \cdots, y_{Ns}]^T$, $\mathbf{f}_s^H = [1, e^{-j\frac{2\pi}{M}s}, e^{-j\frac{2\pi}{M}2s}, \cdots, e^{-j\frac{2\pi}{M}(L-1)s}]$ and $\tilde{\mathbf{w}}_s = [\tilde{w}_{1s}, \tilde{w}_{2s}, \cdots, \tilde{w}_{Ns}]^T$. It is assumed that each entry of $\tilde{\mathbf{w}}_s$ is an i.i.d ZMCSCG random variable with variance $\tilde{\sigma}^2$. Now, let us assume that we are interested in estimating $d_{ks}$ using a beamforming vector $\mathbf{a}_{ks} \in \mathcal{C}^{N \times 1}$ as

$$\hat{d}_{ks} = \mathbf{a}_{ks}^H \mathbf{y}_s = \mathbf{a}_{ks}^H (\sum_{i \neq k}^{K} \mathbf{H}_i^T \mathbf{f}_s^* d_{is} + \tilde{\mathbf{w}}_s). \tag{5}$$

As can be seen from this expression, there are two main independent components: $\mathbf{a}_{ks}^H \mathbf{H}_i^T \mathbf{f}_s^*$ and $\mathbf{a}_{ks}^H \tilde{\mathbf{w}}_s$, and the average powers of these terms are given as (see Appendix A of [16])

$$\begin{aligned}
\mathrm{E}\{|\mathbf{a}_{ks}^H \mathbf{H}_i^T \mathbf{f}_s^*|^2\} =& \overbrace{|\mathbf{t}_{ks}^H \mathbf{X}_i^* \tilde{\mathbf{F}}^T \mathbf{C}_{ii} \mathbf{f}_s^*|^2}^{\triangleq \theta_{kis}} \\
&+ \overbrace{\sum_{m \neq i}^{K} \mathbf{t}_{ks}^H \mathbf{X}_m^* \tilde{\mathbf{F}}^T \mathbf{C}_{imms} \tilde{\mathbf{F}}^* \mathbf{X}_m^T \mathbf{t}_{ks}}^{\triangleq \mu_{kis}} \\
&+ \overbrace{\mathbf{t}_{ks}^H \left( \sum_{m=1}^{K} \sum_{u \neq m}^{K} \mathbf{X}_u^* \tilde{\mathbf{F}}^T \mathbf{C}_{imus} \tilde{\mathbf{F}}^* \mathbf{X}_u^T \right) \mathbf{t}_{ks}}^{\triangleq \bar{\mu}_{kis}} \\
\mathrm{E}\{|\mathbf{a}_{ks}^H \tilde{\mathbf{w}}_s|^2\} =& \sigma^2 \overbrace{\sum_{i=1}^{K} \mathbf{t}_{ks}^H \mathbf{X}_i^* \tilde{\mathbf{F}}^T \mathbf{C}_{ii} \tilde{\mathbf{F}}^* \mathbf{X}_i^T \mathbf{t}_{ks}}^{\triangleq \xi_{ks}} \\
&+ \overbrace{\mathbf{t}_{ks}^H \left( \sum_{i=1}^{K} \sum_{m \neq i}^{K} \mathbf{X}_i^* \tilde{\mathbf{F}}^T \mathbf{C}_{im} \tilde{\mathbf{F}}^* \mathbf{X}_m^T \right) \mathbf{t}_{ks}}^{\triangleq \bar{\xi}_{ks}}
\end{aligned} \tag{6}$$

where $\mathbf{C}_{ki} = \mathrm{E}\{\mathbf{H}_k^* \mathbf{H}_i^T\}$ and $\mathbf{C}_{kims} = \mathrm{E}\{\mathbf{H}_i^* \mathbf{H}_k^T \mathbf{f}_s^* \mathbf{f}_s^T \mathbf{H}_k^* \mathbf{H}_m^T\}$. With the proposed approach, one can achieve the following average SINR with $\hat{d}_{ks}$

$$\bar{\gamma}_{ks} = \frac{\mathrm{E}\{|\mathbf{a}_{ks}^H \mathbf{H}_k^T \mathbf{f}_s^*|^2\}}{\sum_{i\neq k}^K \mathrm{E}\{|\mathbf{a}_{ks}^H \mathbf{H}_i^T \mathbf{f}_s^*|^2\} + \mathrm{E}\{|\mathbf{a}_{ks}^H \tilde{\mathbf{w}}_s|^2\}}$$
$$= \frac{\theta_{kks} + \mu_{kks} + \bar{\mu}_{kks}}{\sum_{i\neq k}^K (\theta_{kis} + \mu_{kis} + \bar{\mu}_{kis}) + \xi_{ks} + \bar{\xi}_{ks}}. \quad (7)$$

## V. Proposed two phase scheduling design

Once the joint CSI estimation and beamforming procedures are determined, the next phase will be to perform UE scheduling. In general, a scheduler is designed to optimize some performance criteria. In the current paper, we consider the maximization of the total sum SE achieved by all sub-carriers with a per UE power constraint since it is an uplink transmission. This problem can be mathematically formulated as

$$\max_{K,\mathbb{K},\mathbf{t}_{ks}} \sum_{s=1}^{M} \sum_{k=1}^{K} \log(1 + \bar{\gamma}_{ks}), \; P_{ks} \leq P_{max}, \quad k \in \mathbb{K}_{max} \quad (8)$$

where $P_{max}$ is the maximum transmitted power at each sub-carrier of each UE and $\mathbb{K}$ is the scheduled set of UEs out of $\mathbb{K}_{max}$ sets. This problem is a mixed integer programming non-convex problem whose global optimal solution is difficult to obtain. In the following, we discuss the proposed two phase approach to solve this problem sub-optimally.

For the appropriately selected $K$ and $\mathbf{x}_k$, one can exploit the following behaviors in (6): $\theta_{kis}$ scales with $N^2$, $\mu_{kis}$ and $\xi_{ks}$ scale with $NL$, and $\bar{\mu}_{kis}$ and $\bar{\xi}_{ks}$ decreases as $N$ increases. For this reason, we propose a two stage scheduling approach. In the first stage, we employ $\theta_{kis}$ to carefully select the maximum number of UEs $K_{max}$ and $\mathbf{x}_k$ while ensuring that $\bar{\gamma}_{ks}$ with $N$ and becomes unbounded as $N \rightarrow \infty$. This is due to the fact that $N \gg L$ in a massive MIMO regime for a fixed $L$. In the second phase, we tune $1 \leq K \leq K_{max}$ and select the corresponding UEs using greedy based scheduling by taking into account $\theta_{kis}$, $\mu_{kis}$ and $\xi_{ks}$ as the contributions of these terms scale with $LN$.

In the first phase, we select $K_{max}$ to maintain a non-decreasing $\bar{\gamma}_{ks}$ with $N$ and becomes unbounded as $N \rightarrow \infty$. In fact one can express $\bar{\gamma}_{ks}$

$$\bar{\gamma}_{ks} = \frac{\frac{1}{N}\theta_{kks} + \beta_0}{\sum_{i\neq k}^K \frac{1}{N}\theta_{kis} + \beta_1} \quad (9)$$

where $\beta_0 = \frac{1}{N}(\mu_{kks} + \bar{\mu}_{kks})$ and $\beta_1 = \frac{1}{N}(\sum_{i\neq k}^K(\mu_{kis} + \bar{\mu}_{kis}) + \xi_{ks} + \bar{\xi}_{ks})$ which are the terms independent of $N$. It follows that to maintain a non-decreasing $\bar{\gamma}_{ks}$ with $N$ and becomes unbounded as $N \rightarrow \infty$, $\theta_{kks}$ should increase with $N^2$ while maintaining $\theta_{kis} = 0, i \neq k$. One straightforward way of achieving this is by designing $\mathbf{x}_k$ and $\mathbf{t}_{ks}$ while ensuring

$$\max_{K,\mathbf{x}_k,\mathbf{t}_{ks}} |\mathbf{t}_{ks}^H \bar{\mathbf{F}}_{ks} \mathbf{x}_k^*|$$
$$\text{s.t } \mathbf{t}_{ks}^H \bar{\mathbf{F}}_{is} \mathbf{x}_i^* = 0, \forall i \neq k \quad (10)$$

where $\bar{\mathbf{F}}_{ks} = \mathrm{diag}(\tilde{\mathbf{F}}^T \mathbf{C}_{kk} \mathbf{f}_s^*)$. For arbitrary $\mathbf{X}_k$ and $\mathbf{C}_{kk}$, a linear combination vector $\mathbf{t}_{ks}$ ensuring the equality constraints does not exist if $K > \tilde{M}$. This can be verified by considering i.i.d channel $\mathbf{C}_{kk} = \mathbf{I}$. For this reason, we allow a maximum of $K_{max} = \tilde{M}$ UEs to transmit their pilots at the pilot transmission phase[2]. In fact, one can maintain an increasing average SINR with increasing $N$ if and only if $|\mathbf{t}_{ks}^H \bar{\mathbf{F}}_{ks} \mathbf{x}_k^*| > 0, \mathbf{t}_{ks}^H \bar{\mathbf{F}}_{is} \mathbf{x}_i^* = 0, \forall i \neq k$ is maintained. This is possible when we select $\mathbf{x}_k$ which ensures $\mathbf{Q}_s \triangleq [\bar{\mathbf{F}}_{1s}\mathbf{x}_1^*, \bar{\mathbf{F}}_{2s}\mathbf{x}_2^*, \cdots, \bar{\mathbf{F}}_{Ks}\mathbf{x}_K^*]$ is a full rank matrix. For a general setup, it is not trivial to select $\mathbf{x}_k$ to ensure a full rank $\mathbf{Q}_s$ for any $\mathbf{C}_{kk}$ and $\tilde{M} = M$ (if it exists). However, we have observed that $\mathbf{Q}_s, \forall s$ achieves higher rank when $\mathbf{x}_k, \forall k$ are selected from the orthogonal Zadoff-Chu sequence. This motivates us to select $\mathbf{x}_k, \forall k$ as the $K$ orthogonal Zadoff-Chu sequences.

In the second phase, we select $1 \leq K \leq K_{max}$ by taking into account $\theta_{kis}$, $\mu_{kis}$ and $\xi_{ks}$ as the contributions of these terms scale with $LN$. From the first phase, although $K = K_{max}$ UEs can be scheduled for data transmission such a choice of $K$ may not yield the maximum sum SE (9)[3]. Thus, for a given $N$, SNR and channel environment, we search the appropriate $K$ exhaustively that yields the maximum total sum SE over all sub-carriers. One way of selecting $K$ UEs is by considering all possible combinations of the UEs. However, such an approach has an exponential complexity. For this reason, we incrementally set $K$ and select the corresponding UEs by applying the well known greedy based scheduling approach to significantly reduce the design complexity [18]. We would like to mention here that one can also further improve the total sum SE by greedily selecting the UEs per sub-carrier from the $K$ UEs. More details in this direction can be found in [16].

We would like to mention here that with the existing approach, we have obtained $\tilde{K}$. One can also apply greedy based approach to schedule only a subset of the UEs while maximizing the total sum rate. Such an optimization, however, may need to be performed for each channel realization which makes this approach computationally expensive compared to that of the proposed approach which uses channel statistics to schedule the UEs. This is due to the fact that the proposed scheduling is performed when the channel covariance matrix is modified which occurs in a much slower rate compared to the CSI variation.

## VI. Simulation Results

This section provides simulation results. In general, different multipath channel propagation models are suggested for different standards, where each model characterizes both the small scale and large scale fadings. In a typical environment, the large scale fading captures shadowing, power delay profile (PDP), and distance dependent path loss, spatial correlation over multiple antennas and temporal correlation over multipath channel taps [19]. In this simulation, we ignore the effect of shadowing, Doppler effect, and assume that all UEs experience the same distance dependent path loss and PDP. Further, we utilize the PDP information to get the channel coefficients by assuming that the small scale fading corresponding to each delay tap is modeled by a Rayleigh fading. With these settings, $\bar{h}_{knl}, l = 1, \cdots, L$ becomes

$$\bar{h}_{knl} = \sum_{i=1}^{\tilde{L}} \bar{q}_i \bar{\bar{h}}_{kni} g((l-Q)T_s - \tau_i),\ l = 1, \cdots, L \quad (11)$$

where $\tilde{L}$ is the size of PDP coefficients, $Q$ is a shift parameter selected to maintain $\bar{h}_{knl} = 0, l \leq 0$, $g(t)$ is the overall transceiver pulse shaping filter, $\tau_i$ is the delay coefficient corresponding to the $i$th path, and $\bar{\bar{h}}_{kni}$ is the Rayleigh fading coefficient corresponding to the $i$th path, $k$th UE and $n$th BS antenna. One can observe from (11) that $L$ depends on several factors including $g(t)$, $T_s$ and $\tau_i$. In fact, for a given $T_d = \max_{i,j} |\tau_i - \tau_j|$, different sets of $\tau_i, \bar{q}_i$ may lead to a different statistical properties of $\bar{h}_{knl}$.

[2]Note that in some scenario, it is possible to allow $K > \tilde{M}$ (for example when $\mathbf{C}_{ii}\mathbf{C}_{kk} = \mathbf{0}, \exists i$ as discussed in [5] and [17]). However, such a special case is not considered here.

[3]This will be demonstrated in the simulation section.

All results are obtained by averaging 10000 channel realizations. The SNR is defined as $\gamma = \frac{\mathrm{E}\{|d_i|^2\}}{\sigma^2}$. We set $\tilde{K} = 4$ and $\mathbf{R}_k = \mathbf{I}$, the temporal correlation matrices of multipath channels are computed from the PDP, and $g(t)$ is set as a raised cosine filter (RCF) with roll off factor $\alpha = 0.25$[4].

### A. *Effect of PDP and bandwidth on* $\mathrm{rank}(\mathbf{Q}_s)$

The performance of the scheduler depends on $\mathrm{rank}(\mathbf{Q}_s)$, and $L$ generally increases with $B$ for a fixed delay spread $T_d$ and $g(t)$ which ultimately help increase $K_{max}$. In this subsection, we examine $\mathrm{rank}(\mathbf{Q}_s)$ for dense and sparse channel paths. In general, one may experience $\mathbf{C}_{kk}$ be closer to identity in the dense channel paths which yields $\mathrm{rank}(\mathbf{Q}_s) \approx K_{max}$. On the other hand, for the case of sparse channel paths, $\mathrm{rank}(\mathbf{Q}_s)$ will likely be much less than $K_{max}$.

To demonstrate this, we plot the CDF of the SVD of $\mathbf{Q}_s$ (i.e., the percentage of $\mathrm{SVD}(\mathbf{Q}_s) > Th$) averaged over all sub-carriers and UEs for different threshold values $Th$ when the temporal covariance matrices of all UEs are computed from the PDP parameters of the LTE and custom designed PDPs as shown in Fig. 3. One can observe from this figure that the the percentage of $\mathrm{SVD}(\mathbf{Q}_s) > Th$ increases as the multipath density increases. This can be confirmed by comparing the $\mathrm{CDF}(\mathrm{SVD}(\mathbf{Q}_s) > Th)$ of uniformly spaced delays with $\Delta\tau = 10$ and $\Delta\tau = 50ns$, and the LTE environments LTE-EPA, LTE-ETA, and LTE-ETU. Next we demonstrate the size of $\mathrm{SVD}(\mathbf{Q}_s) > Th$ averaged over all sub-carriers and UEs for different values of $B$ as shown in Fig. 4. As can be seen from this figure, increasing $B$ helps increase $L$ and the increment has a significant performance gain when $\tilde{L}$ is large which is expected.

The figures 3 and 4 assume that the total number of UEs sending their pilot is set to $K = K_{max} = \tilde{M} = \tilde{K}L$. For this setup, in fact, $\mathrm{CDF}(\mathrm{SVD}(\mathbf{Q}_s) > Th)$ decreases as $Th$ increases in all environments which may not be desirable for a high SNR transmission applications. For arbitrary $Th$, one way of increasing the CDF of $\mathrm{SVD}(\mathbf{Q}_s) > Th$ is by decreasing $K$ as desired. Fig. 5 shows the CDF of $\mathrm{SVD}(\mathbf{Q}_s) > Th$ for different values of $Th$ and $T_s = 7.68\mu s$ (i.e., $B$ = 5MHz, $M = 256$) for LTE-ETU and uniformly spaced delay setup with $\Delta\tau = 10$ns. One can observe from this figure and Fig. 3 that decreasing $K$ definitely helps increase the number of UEs experiencing higher SNR.

[4]Note that one can obtain $g(t)$ by employing a square root RCF with $\alpha = 0.25$ at the transmitter and receiver.

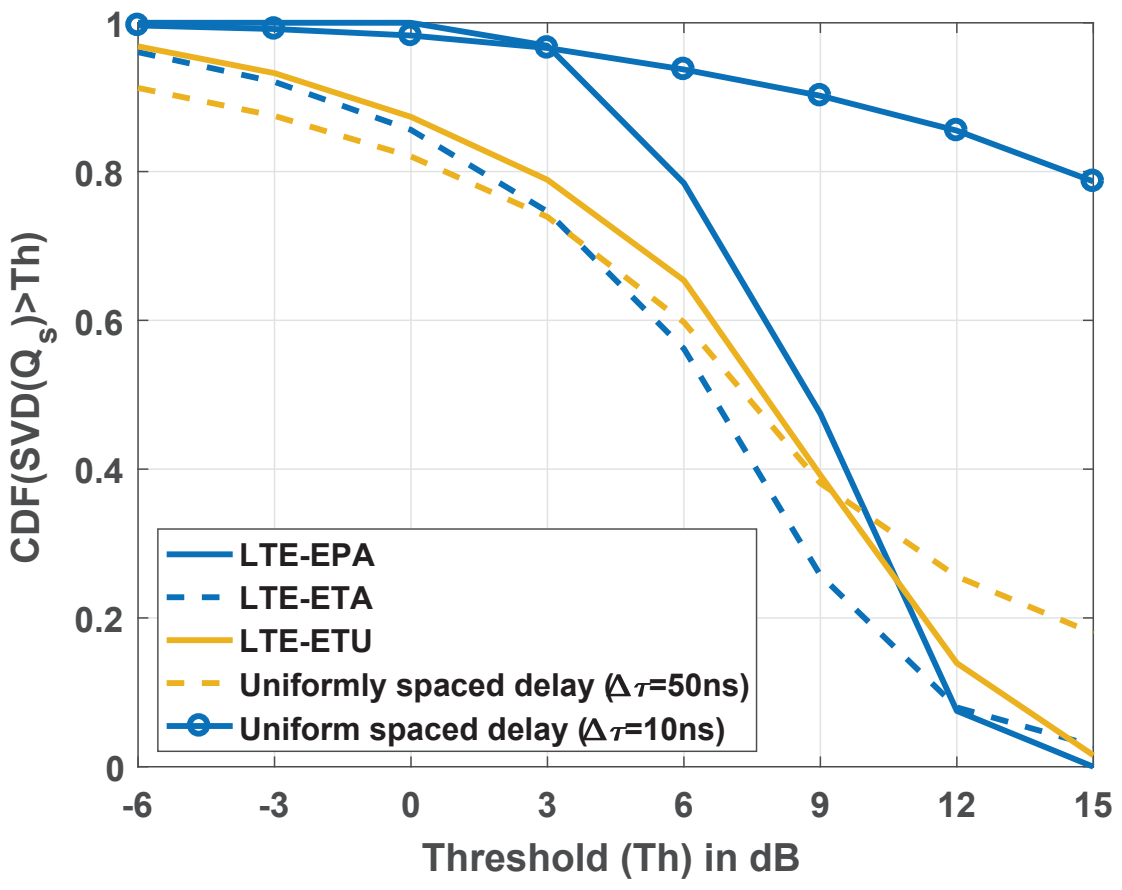

Fig. 3: The CDF of the $\mathrm{SVD}(\mathbf{Q}_s) > Th$ averaged over all sub-carriers and UEs when $T_s = 7.68\mu s$ (i.e., $B$ = 5MHz, $M = 256$) and different environments. For the uniformly spaced delay setup, we use $\bar{q}_i = 0$ and $[\tau_1, \tau_2, \cdots] = [0 : \Delta\tau : 5000]$.

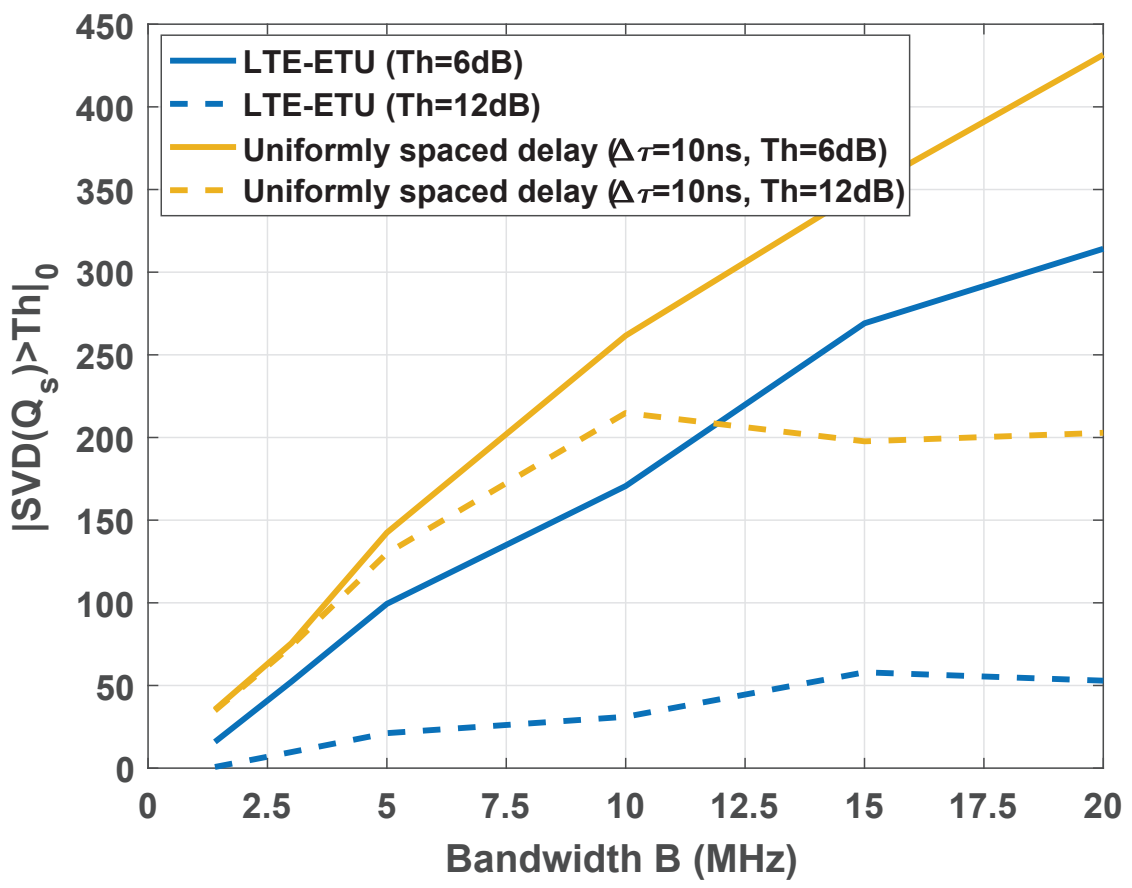

Fig. 4: The $|\mathrm{SVD}(\mathbf{Q}_s) > Th|_0$ averaged over all sub-carriers and UEs for different values of $B$ with LTE-ETU and uniformly spaced delay parameter setups.

### B. *Comparison of OFDM and proposed design*

In this simulation, we compare the SE performance of the proposed channel estimation, beamforming and two phase scheduling design with that of the conventional OFDM approach. In this regard, we consider the PDP's used in LTE-ETU environment with $B$ = 5MHz, $SNR$ = 0dB and different values of $N$ and $K$ as shown in Fig. 6. One can observe from this figure that different choices of $K$ yields different average spectrum efficiency (SE). Furthermore, when $K$ increases the proposed approach requires large $N$ to achieve better total SE compared to the conventional OFDM approach. This validates that the benefit of the proposed design can be exploited by tuning $K$ according to the channel environment and number of BS antennas.

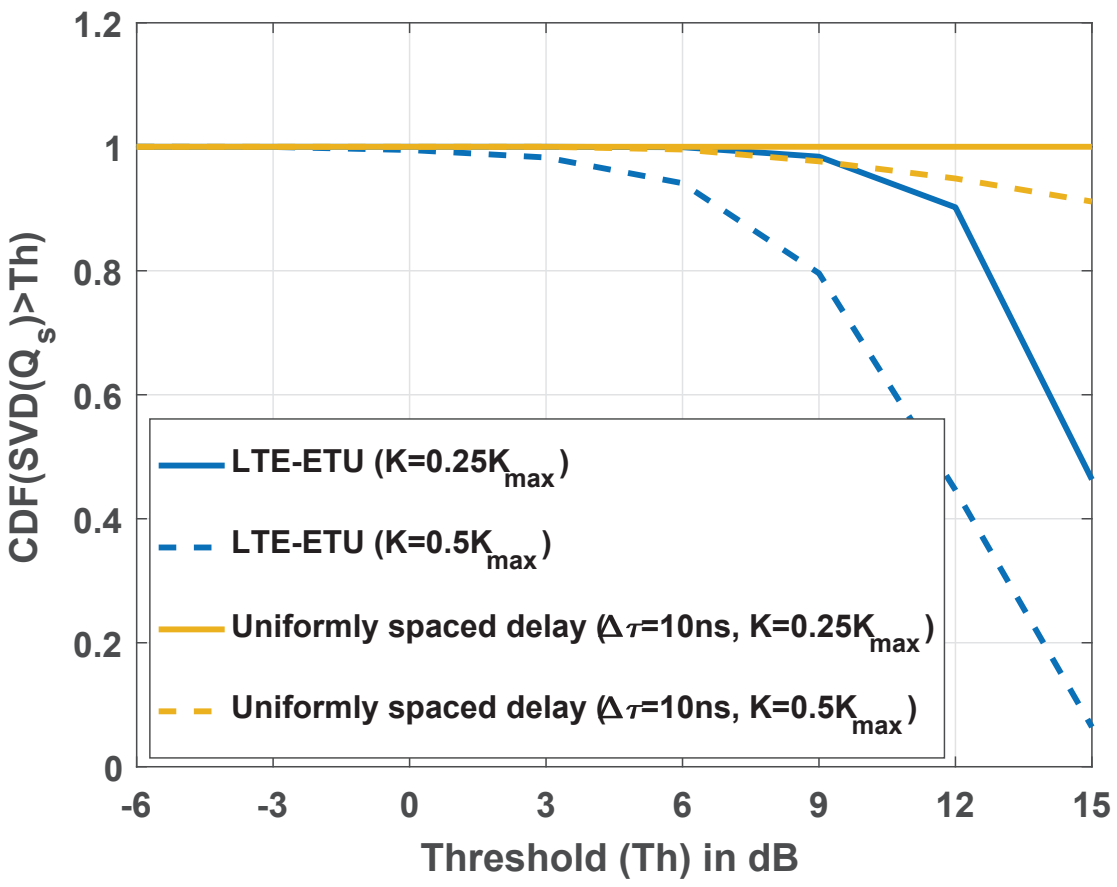

Fig. 5: The CDF of the $\mathrm{SVD}(\mathbf{Q}_s) > Th$ averaged over all sub-carriers and UEs when $T_s = 7.68\mu$s (i.e., $B = 5$MHz, $M = 256$) for different values of $K$. For the uniformly spaced delay setup, we use $\bar{q}_i = 0$ and $[\tau_1, \tau_2, \cdots] = [0 : \Delta\tau : 5000]$.

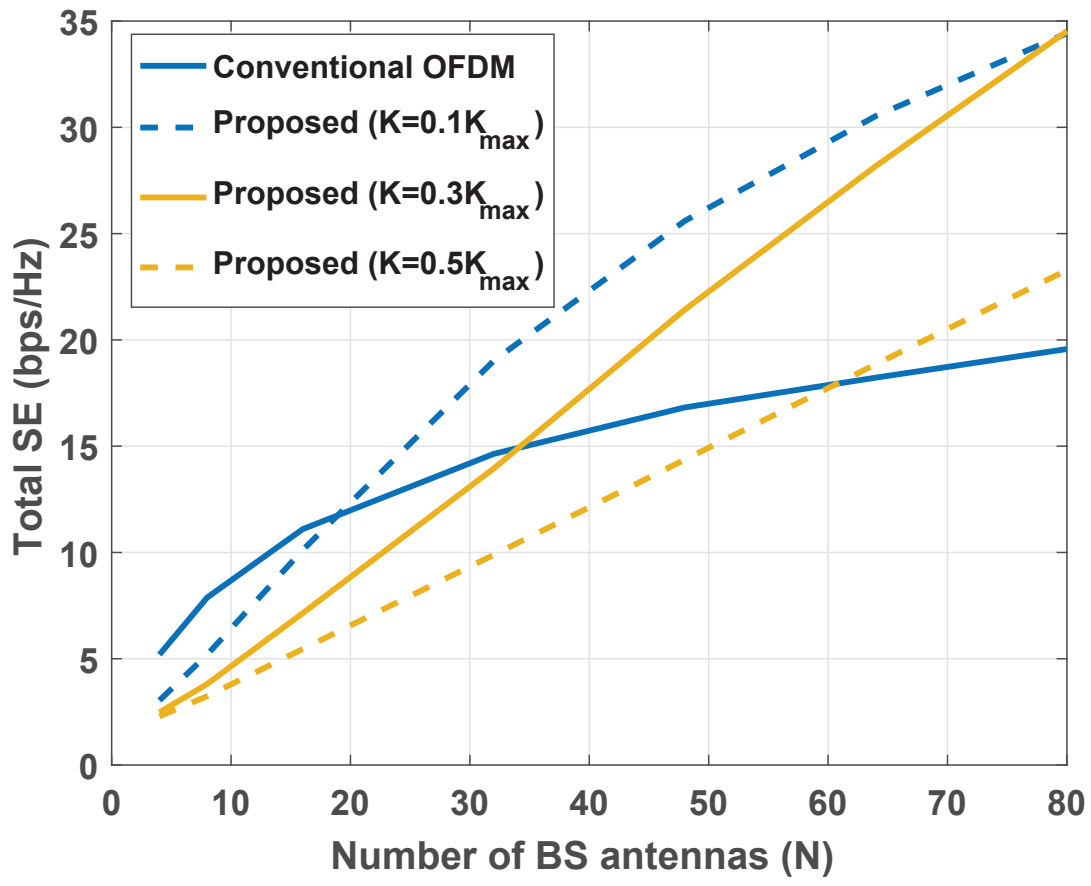

Fig. 6: Comparison of OFDM and proposed design when $T_s = 7.68\mu$s (i.e., $B = 5$MHz, $M = 256$) for different values of $K$ in the LTE-ETU environment.

## VII. Conclusions

This paper proposes a novel approach for designing channel estimation, beamforming and scheduling jointly for wideband massive MIMO systems. With the proposed approach, we first quantify the maximum number of UEs that can send pilots which may or may not be orthogonal. Specifically, when the channel has a maximum of $L$ multipath taps, and allocate $\tilde{M}$ sub-carriers for the CSI estimation, a maximum of $\tilde{M}$ UEs' CSI can be estimated (i.e., $L$ times larger compared to the traditional MMSE based CSI estimation approach) in a massive MIMO regime. Then, we select only a subset of these UEs $K$ for data transmission using the greedy based scheduling for each sub-carrier with the proposed joint beamforming and scheduling design. All the analytical expressions have been demonstrated via numerical results, and the superiority of the proposed design over the conventional transmission approach is demonstrated using extensive numerical simulations for LTE channel models. In particular, for larger $K$, the proposed approach requires larger $N$ to achieve better total SE compared to the conventional OFDM approach. Consequently, the benefit of the proposed design can be exploited by tuning $K$ according to the channel environment and number of BS antennas.